\documentclass[11pt]{article}
\usepackage{amsfonts}
\def\be{\begin{eqnarray}}
\def\ee{\end{eqnarray}}
\def\*{\star}
\begin{document}
          \begin{flushright}  
          ANL-HEP-PR-01-111 \\ Miami TH/2/01 \\  RU-01-18-B
           \end{flushright}

{\LARGE
\centerline{Area Potentials and Deformation Quantization} }
\centerline {\phantom{aaa}} 
  {\large
\centerline{Thomas L Curtright$^{\S}$, Alexios P 
Polychronakos$^{\natural}$\footnote{
On leave from the Theoretical Physics Department, Uppsala University, Sweden.}, 
 and Cosmas K Zachos$^{\P}$}  }
$^{\S}$ Department of Physics, University of Miami,
Box 248046, Coral Gables, Florida 33124, USA\\
\phantom{a} \qquad\qquad{\sl curtright@physics.miami.edu}  

$^{\natural}$  Physics Department, The Rockefeller University,
 New York, NY 10021, USA\\
\centerline{and} \\
\phantom{AaA} Physics Department, University of Ioannina, 
45110 Ioannina, Greece\\
\phantom{a} \qquad\qquad{\sl poly@teorfys.uu.se}  

$^{\P}$ High Energy Physics Division,
Argonne National Laboratory, Argonne, IL 60439-4815, USA \\
\phantom{a} \qquad\qquad{\sl zachos@hep.anl.gov} 

\begin{abstract} 
Systems built out of N-body interactions, beyond 2-body interactions, are 
formulated on the plane, and investigated classically and quantum mechanically 
(in phase space). Their Wigner Functions---the density matrices in phase-space 
quantization---are given and analyzed. 
\end{abstract}
\noindent\rule{7in}{0.02in}

In this brief note, we consider systems of $N$ particles on a plane, 
interacting through N-body potentials, over and above more familiar 2-body 
potentials.  
Below, we provide the compact generating function, 
\begin{equation}  
G({\bf a,a^*,b,b^*;z,z^*,p,p^*} ) = 
\frac{1}{\pi^{2N}} \exp \frac{1}{\hbar} \left (
{ |{\bf a}|^2 + |{\bf b}|^2  \over 2 }  
-m ({\bf  z}- {\bf a} )^*  
\cdot {\mathfrak W} ({\bf  z}-  {\bf a} )  
-{ ({\bf p} -{\bf b} )^*  \cdot ({\bf p} -{\bf b} ) \over m}  \right ), 
\end{equation} 
for their Wigner Functions (WF), the density matrices in 
phase-space quantization
\cite{moyal,wigner,weyl,bayen}. (For reviews, see \cite{reviews}.) 

Consider the hamiltonian of three particles on a 
plane, interacting in proportion to the (signed) {\em area} of the triangle
they define:
\begin{equation} 
H_3={ {\bf p}_1^2+{\bf p}_2^2+{\bf p}_3^2 \over 2m} +m A  ~,
\end{equation}
where the signed area (doubled) is 
\begin{equation} 
A= ({\bf r}_1 -{\bf r}_2)\wedge ({\bf r}_2-{\bf r}_3) .
\end{equation}

Its (saddle) quadratic form, with the three 2-vectors arrayed in succession,
 ${\bf R}= (x_1,y_1,x_2,y_2,x_3,y_3)$, is 
\begin{equation} 
\left( \begin{array}{cccccc}
0&0&0&1&0&-1\\
0&0&-1&0&1&0\\
0&-1&0&0&0&1\\
1&0&0&0&-1&0\\
0&1&0&-1&0&0\\
-1&0&1&0&0&0\\
\end{array} \right) ~= 
\left( \begin{array}{ccc}
0&1&-1\\
-1&0&1\\
1&-1&0\\
\end{array} \right) \otimes 
\left( \begin{array}{cc}
0&1\\ 
-1&0\\
\end{array} \right)~.
\end{equation}

This may be simplified by elimination of the $2\times 2$ structure.
On the plane, rotational invariance 
can be exploited by introducing complex variables, $z\equiv x+iy$, which 
cuts down the size of such matrices by half on each side, since
\begin{equation} 
{\bf r}_1 \wedge {\bf r}_2= x_1 y_2 - x_2 y_1= {-i\over 2}~ (z_1^*, z_2^*) 
\left( \begin{array}{cc}
0&1\\
-1&0\\
\end{array} \right) ~ \left( \begin{array}{c}
z_1\\
z_2\\
\end{array} \right) ~.
\end{equation}
Clearly multiplication of the $z$s by an arbitrary phase yields a rotated
configuration with the same eigenvalue (a double degeneracy of the 
original problem).

Thus the above quadratic form hermitean matrix (up to 1/2) reduces to the 
imaginary antisymmetric matrix
\begin{equation} 
{\bf \Omega}= -i\left( \begin{array}{ccc}
0&1&-1\\
-1&0&1\\
1&-1&0\\
\end{array} \right) .
\end{equation}
Consequently, the above hamiltonian amounts to 
\begin{equation} 
H_3={1\over 2m} ~ {\bf p}^* \cdot {\bf p}  + 
{m\over 2}~ {\bf z}^* \cdot {\bf \Omega  z}~,
\end{equation}
for complex 3-vectors ${\bf z}$ and their canonical conjugates ${\bf p}^*$ 
(with components $p\equiv p_x+ip_y$). 

The matrix ${\bf \Omega}$ consists of Sylvester's 
celebrated ``nonions" \cite{sylvester},
today's standard clock and shift matrix basis. Specifically, 
${\bf \Omega}= -i (h- h^2)$, where 
$h$ is the cyclic permutation shift matrix, 
with $h^3= {1\kern-0.35em\llap~1}$,
hence $h^t= h^{-1} =h^2$. 
For $\omega=- 1/2+ i\sqrt{3} /2$,
a cube root of unity, ${\bf \Omega}$ is diagonalized by the 
Finite Fourier Transform 
unitary matrix \cite{fft},
\begin{equation} 
U= \frac{1}{\sqrt{3}} \left( \begin{array}{ccc}
1&\omega & \omega ^2     \\
1&\omega ^2 &\omega \\
1&1&1\\
\end{array} \right) .
\end{equation}
\begin{equation} 
U^\dagger \Omega U = -i (\omega - \omega ^2)  
\left( \begin{array}{ccc}
0&0 & 0 \\
0& 1 & 0\\
0&0&-1\\
\end{array} \right) ,
\end{equation}
i.e., the (real) difference of two clock matrices.

The normal mode frequencies-squared of the hamiltonian then are 
0, $\sqrt{3}$, and $-\sqrt{3}$.
The negative eigenvalues (imaginary frequencies) 
reflect the cyclic symmetry, $1,2,3$, 
anti-mirrored into the antistandard order, $1,3,2$. For every eigenvalue,
the opposite eigenvalue follows for the 
two-particle permuted (here, antistandard) configuration.

The corresponding eigenvectors describe the translation zero mode,  
$( 1,  1, 1)$, with all three particles moving in coincidence, so without 
a force;  the pulsating cube roots of unity configuration in 
standard, 1,2,3,  order, $(\omega , \omega ^2 , ~1)$, 
\begin{picture}(100,100)  \thicklines
\put(50,50){\line(1,0){27}}
\put(50,50){\line(-3,5){16}}
\put(50,50){\line(-3,-5){16}}
\put(90,50){\makebox(0,0)[cc]{{\bf r}$_3$}} 
\put(20,80){\makebox(0,0)[cc]{{\bf r}$_1$}} 
\put(20,20){\makebox(0,0)[cc]{{\bf r}$_2$}} 
\end{picture}
, or any rotation of it on the plane (with which it can be combined to 
rotating configurations); and, finally, its mirror image (c.c.),
$( \omega^2 , \omega , ~1)$,
in antistandard, 1,3,2, order. These last (imaginary frequency)
unstable modes lead to the indefinite growing of the area of the 
antistandard-order triangle, corresponding to a quantum mechanical spectrum 
unbounded below.

The  unstable normal modes of increasing negative area and the concomitant
lower unboundedness of the spectrum may be counteracted by considering 
additional harmonic two-body interactions between each pair of particles, 
i.e., a positive semi-definite potential component,
  \begin{equation} 
V_2=\frac{m}{2\sqrt{3}} \left ( 
({\bf r}_1 -{\bf r}_2)^2+  ({\bf r}_2 -{\bf r}_3)^2+({\bf r}_3 -{\bf r}_1)^2
\right ) , 
 \end{equation}
whose corresponding quadratic form matrix,
\begin{equation} 
{\bf W}= \frac{1}{\sqrt{3}}  \left( \begin{array}{ccc}
2& -1& -1\\
-1& 2& -1\\
-1& -1& 2\\
\end{array} \right) = \frac{1}{\sqrt{3}} \left (  
3~ {1\kern-0.35em\llap~1} -\left( \begin{array}{ccc}
1& 1& 1\\
1& 1& 1\\
1& 1& 1\\
\end{array} \right) \right ),
\end{equation}
projects out the translation mode and has eigenvalue $\sqrt{3}$ for the other 
modes. Thus, the potential $A+V_2 $, corresponds to the modified form 
${\mathfrak W }={\bf \Omega}+{\bf W}$, 
whose eigenvalues are thus shifted from the above ones to non-negative ones, 
0, $2\sqrt{3}$, and 0, for the same orthogonal eigenvectors.
Consequently, considering the plane-rotational invariance doubling of the 
modes, the system ${\mathfrak H}_3$ resolves to two harmonic oscillators 
and four free modes, whose quantization yields a 2-d linear spectrum on the 
free particles' continuum.

The system considered so far naturally generalizes to a  
standard (anticlockwise ordered) $N$-gon, with (twice) the area, 
consisting of a fan of triangles,
\begin{equation} 
A= {\bf r}_1\wedge {\bf r}_2+
{\bf r}_2\wedge {\bf r}_3+...+
{\bf r}_N\wedge {\bf r}_1.
\end{equation} 
It corresponds to the $N\times N$ matrix \cite{sylvester,weyl,fz}
\begin{equation}
{\bf \Omega} =-i ( h - h^{N-1}). ...=2 \sin \left ( \ln (-ih) \right ) ,
 \end{equation}
where $h$ is the $N\times N$ cyclic shift matrix, 
with $h^N= {1\kern-0.35em\llap~1}$ and $h^t= h^{-1} =h^{N-1}$. Extension 
from $N$ to $N+1$ amounts to the addition of just one triangle to the fan.

Thus, the general eigenvalues for ${\bf \Omega}$ are 
\begin{equation}
\lambda_k=-i ( \omega^k - \omega^{-k}),
 \end{equation}
with $\omega$ the primary $N$-th root of unity and $k=0,1,2,...,N-1$.
Note the evident mirror-image pairing of the eigenvalues
$\lambda_k=-\lambda_{N-k}$.  
The corresponding $N$ complex eigenvectors are
\begin{equation}
| e_k\rangle =(1,  \omega^k,  \omega^{2k},  ...,\omega^{k(N-1)}),
 \end{equation}
which array into the corresponding unitary diagonalizing 
Finite Fourier Transform matrix $U$.

The negative eigenvalues may be shifted, as above,  
by the addition of a 2-body interaction in the potential,
  \begin{equation} 
V_2=\frac{m}{2c} \sum_{i<j}  ({\bf r}_i -{\bf r}_j)^2= 
\frac{m}{2c} \left ( N\sum_{i}  {\bf r}_i^2- (\sum_{i}  {\bf r}_i)^2 \right ),
 \end{equation}
where $c$ is a normalization constant. This interaction corresponds 
to the matrix 
  \begin{equation} 
{\bf W}=\frac{1}{c} ( N~~{1\kern-0.35em\llap~1} -|e_0\rangle \langle e_0 | ). 
\end{equation}
$|e_0\rangle =(1,1,1,1,...,1)$ is the translation zero mode projected out, 
while all other eigenvalues but its own are shifted by $N/c$. Thus, choosing 
the normalization ($c\leq N/2$ will suffice), all negative eigenvalues 
of ${\mathfrak W} $ may be neutralized to yield a stable system 
${\mathfrak H}_N$ with a spectrum bounded below.

The classical equations of motion are
\begin{equation} 
\ddot{\bf z} = - {\bf \Omega z}.   \label{EL}
\end{equation} 
Note the vanishing of the torque (and hence 
conservation of angular momentum) for the system 
$H_N$ (manifestly rotational invariant!), even though the interaction 
is not two-body. This follows from 
summing the torque on each particle, since ${\bf \Omega}$ is hermitean,
\begin{equation} 
\tau= \sum_i {\bf r}_i \wedge m \ddot{\bf r}_i  
={im \over 2} ( {\bf z}^* \cdot {\bf \Omega z} -  ({\bf \Omega  z})^* \cdot 
{\bf z} )=0.
\end{equation} 

Further note the $N$ independent time-invariant complex vector 
combinations,
\begin{equation} 
\exp (-i \sqrt{ {\bf \Omega}} ~ t) ~~\left  ( 
{\bf p} + i \sqrt{ {\bf \Omega}}~ {\bf z} \right ) .
\end{equation} 
In the less compact notation of $2n$-dimensional phase space ($n=2N$) 
there are $2n$ real independent time-invariant combinations.
Eliminating $t$ among them yields $2n-1$ independent real time-invariant 
quantities\footnote{ For example, in the normal mode representation of 
${\mathfrak H}_3$ with the full six-dimensional ${\mathfrak W}$ diagonalized to 
diag$(0,0,2\sqrt{3},2\sqrt{3},0,0)$, and the coordinates starting at 0 at
$t=0$, a set of 11 independent invariants is 
$${p_{x_2}+ i2\sqrt{3} x_2   \over p_{y_2}+ i2\sqrt{3} y_2  } ; \qquad 
(p_{x_2}+ i2\sqrt{3} x_2) \exp ( -2\sqrt{3} x_1/m p_{x_1} );  \qquad 
p_{x_1},~ p_{y_1}, ~p_{x_3},~ p_{y_3}; \qquad   
{p_{y_1} x_1 \over p_{x_1}y_1}, ~~ {p_{x_3} x_1  \over p_{x_1}x_3},~~ 
{p_{y_3} x_1 \over p_{x_1}y_3} ~. $$}               
$Q_i$, which  characterize maximally superintegrable 
systems in phase space \cite{superintegrable}. For these, time evolution 
is completely specified by flow perpendicular to {\em all} the phase-space 
gradients $\nabla Q_i$. 
Thus, the time evolution of the phase-space variables, and so the 
time derivative of any phase-space function,
is proportional to the corresponding Jacobian determinant for the full 
$2n$-dimensional phase space, 
\begin{equation} 
{df\over dt}\propto {\partial (f,Q_1,...,...,Q_{2n-1}) \over \partial 
(x_1,p_{x_1},  y_1,p_{y_1},  ...,x_N,p_{x_N}, y_N,p_{y_N})   },
\end{equation} 
i.e., the phase-space Nambu Bracket \cite{superintegrable}.  

There is an alternate way to stabilize the system by shifting the negative 
eigenvalues of ${\bf \Omega}$ without the above introduction of 2-body 
interactions, but, instead, through the introduction of a magnetic 
field $B$ pointing 
into the plane in question, and consideration of the particles as charged.
Such a magnetic field breaks parity and differentiates between left- and 
right-spinning cyclotron orbits. The magnetic field modifies the 
complex equations of   motion (\ref{EL}),
\be
{\ddot {\bf z}} = i B {\dot {\bf z}} - {\bf \Omega} {\bf z}, \label{MLE}
\ee
and thus shifts the eigenfrequencies of the modes from $\sqrt{\lambda_k}$ to
$w_k$, solutions of 
\be
w^2_k - B w_k - \lambda_k = 0.
\ee
The modified mode frequencies
\be
w_k = \frac{B}{2} \pm \sqrt{\frac{B^2}{4} + \lambda_k}
\ee
are real for $B^2 \geq -4\min (\lambda_k )$, which equals  8 for even $N$,
and is smaller for odd $N$. Thus, the system is stable for a field at least
as strong as $|B|>2\sqrt{2}$. (In the small mass limit, where the strength of
the area potential and the magnetic field are kept unchanged, the left-hand 
side of (\ref{MLE}) drops out and hence $w_k=\lambda_k/B$, 
describing pure Landau levels.) 

The zero mode $w_0=0$  coincides with $\sqrt{\lambda_0} =0$
and represents Landau level degeneracy. 
The mode frequencies $w_k$ can be positive or negative,
corresponding to the chirality of the cyclotron orbits. 
At the critical field, $| B_c |=2\sqrt{2}$, (taking 
$-i\omega_k = \exp (2\pi ik /N - i\pi/2 )$), the above square root acts 
on a perfect square, so that  
\be
w_k/2 = \hbox{sgn}(B_c) \cos (\pi/4) \pm  \cos \left ( 
\frac{\pi k}{N} - \frac{\pi}{4} \right ) ,
\ee
where $\hbox{sgn}(B_c) = \pm 1$. So, essentially (for large $N$ where
the vanishing and nonvanishing $w$ corresponding 
to $\lambda_0=0$ are neglected), 
$25\%$ of the modes are of one chirality and $75\%$ are of the opposite.
For the above $N=3$ case, the critical field is 
$B_c=2 (3)^{1/4}$; beyond the 
zero mode, there is one mode with $w=B_c$, two with 
$w= (3)^{1/4} ( 1\pm \sqrt {2})$, and the two ones stabilized by the field 
to $w=B_c/2$.

In the large $N$ limit, the above system reduces to a closed noncovariant 
string undulating on the plane. The particle index becomes a continuous 
periodic variable, $\sigma = 2\pi n/N$, and the complex $N$-vector of
the particles' position goes into a scalar closed string field of that 
variable, ${\bf z}(t) \mapsto \phi (\sigma , t)$. The potential
is the (signed) area enclosed by the string.
The resulting noncovariant lagrangian density is 
\be
{\cal L} = {m\over 2} {\dot {\bar \phi}} {\dot \phi} 
+ i {\bar \phi}\partial_{\sigma} \phi .
\ee
This amounts to Schr\"odinger's Lagrangian, with the roles of space and 
time reversed. The dispersion relation of Schr\"odinger's 
action, $E=p^2/2$, here reverses to 
\be
w^2 + 2k = 0,
\ee
with $k$ the integer Fourier modes of the compact variable $\sigma$. 
This specifies an infinite tower of stable and unstable modes. 
As in the finite case, introducing an additional harmonic two-body
potential for the particles amounts to a term, 
\be
\int_{\sigma < \sigma'} d\sigma d\sigma' |\phi(\sigma) - \phi(\sigma' )|^2 =
\int d\sigma |\phi(\sigma)|^2 - \left| \int d\sigma \phi(\sigma ) \right|^2 ,
\ee
essentially an external harmonic potential minus a center-of-mass term.
This would correspond to an external constant potential in the Schr\"odinger
equation. The modes are shifted by a constant, but still an infinity of
them remain unstable.

We finally quantize the original system in phase-space \cite{reviews}. 
Groenewold's associative $\*$-product \cite{groen},  in our condensed 
notation of complex vectors, 
\begin{equation} 
\star \equiv ~ \hbox{{\Large 
$e^{  i   \hbar(
\stackrel{\leftarrow }{\partial }_{{\bf z}^*} \cdot
\stackrel{\rightarrow }{\partial }_{\bf p} +
\stackrel{\leftarrow }{\partial }_{\bf z}\cdot
\stackrel{\rightarrow }{\partial }_{{\bf p}^*} 
-\stackrel{\leftarrow }{\partial }_{{\bf p}^*}\cdot 
\stackrel{\rightarrow }{\partial }_{\bf z}
-\stackrel{\leftarrow }{\partial }_{\bf p}\cdot
\stackrel{\rightarrow }{\partial }_{{\bf z}^*}     )} $}}~, 
\end{equation} 
is the cornerstone of phase-space quantization. (It allows c-number functions 
in phase space to multiply with each other and with WFs associatively and 
noncommutatively, in perfect parallel to operator manipulations in the 
standard, Hilbert space, formulation of quantum mechanics \cite{reviews}.)
The Poisson Brackets 
which are exponentiated in the $\star$-product are 
\begin{equation} 
\{ z_i^*, p_j  \}= 2 \delta_{ij} ~,   
\end{equation} 
and their complex conjugate, for $i,j=1,...,N$. 

The linear contact transformation $U$ discussed above, which leads to the 
normal mode variables $U^{\dagger} {\bf z}$ and $U^{\dagger} {\bf p}$,
is a linear canonical transformation, and therefore preserves Poisson 
Brackets and $\star$-products \cite{kim}. (To be contrasted to nonlinear 
canonical transformations which actually transform $\star$-products in a 
covariant fashion \cite{cfz}). That is to say, for these transformations 
the $\*$ above is a scalar, i.e. equals to the same 
expression for the  normal mode variables $U^{\dagger} {\bf z}$.  
All formal manipulations below, then, may be equivalently 
conducted in the space of the original or the normal mode variables, 
and no distinction between either variables in $\*$ is necessary. 

Thus, e.g., in terms of the normal modes, the WF for the $N=3$ case 
${\mathfrak H}_3$ is 
a product of the six WFs of each normal mode: four unnormalizable free ones 
and two oscillators of frequency-squared $2\sqrt{3}\equiv w^2$,   
\begin{equation} 
f = \delta({\bf p}_1 - \hbar {\bf k}_1) 
\delta({\bf p}_2 - \hbar {\bf k}_2)  {(-)^{n+s}\over \pi^2} 
\exp \left ( -{1\over\hbar} ({ {\bf p}^2_2\over mw} +mw {\bf r}^2_2 )\right ) 
L_n \left ( {2\over \hbar} ({p_x^2\over   mw} +  mw x_2^2) \right ) 
 L_s \left ({2\over \hbar} ({p_y^2\over mw} + mw y_2^2)\right ),   
\end{equation} 
where the $L_n$'s are Laguerre polynomials with $n,s=0,1,2,....$,
and the two ${\bf k}$s comprise four arbitrary real constants.
These WFs obey the phase-space stargenvalue equation for the complete 
spectrum of the quantum system \cite{cfz},
\begin{equation}  
{\mathfrak H}_3 \* f = \left ( \hbar^2 {{\bf k}_1^2+{\bf k}^2_3\over 2m} +  
\left ( \frac{3}{4}\right ) ^{1/4} \hbar (1+n + s) \right )    f.
\end{equation} 

Nevertheless, the generic $N$ case may be approached in a compact way through 
the WF generating functions of \cite{cuz}, which generalize the WFs for 
coherent states \cite{kim} (the multiparticle generalization for quadratic 
systems may also be found in \cite{dodonov}).
\begin{equation} 
G({\bf a,a^*,b,b^*;z,z^*,p,p^*} ) = 
\frac{1}{\pi^{2N}} \exp \frac{1}{\hbar} \left (
{ |{\bf a}|^2 + |{\bf b}|^2  \over 2 }  
-m ({\bf  z}- {\bf a} )^*  
\cdot {\mathfrak W} ({\bf  z}-  {\bf a} )  
-{ ({\bf p} -{\bf b} )^*  \cdot ({\bf p} -{\bf b} ) \over m}  \right ). 
\end{equation}
For
\begin{equation} 
{\mathfrak H}_N ={1\over 2m} ~ {\bf p}^* \cdot {\bf p}  + 
{m\over 2}~ {\bf z}^* \cdot {\mathfrak W}{\bf  z}~,
\end{equation}
this function satisfies
 \begin{equation} 
{\mathfrak H} \* G \simeq \frac{\hbar}{2}   \left (   2 N + 
({\bf a}-{\bf b})\cdot(\partial_{\bf a} -\partial_{\bf b}) 
+({\bf a}^* -{\bf b}^*)\cdot(\partial_{\bf a^*} -\partial_{\bf b^*}) 
\right ) G ,
\end{equation} 
\begin{equation} 
G \* {\mathfrak H} \simeq \frac{\hbar}{2}   
\left (2  N + 
({\bf a}+{\bf b})\cdot(\partial_{\bf a} +\partial_{\bf b}) 
+({\bf a}^* +{\bf b}^*)\cdot(\partial_{\bf a^*} +\partial_{\bf b^*}) 
\right ) G.
\end{equation} 

The ${\mathfrak W}$ may be diagonalized by unitarily transforming
all variables and shifts ${\bf z, p, a, b}$, $\partial_{{\bf z}^*}$, 
$\partial_{{\bf p}^*}$, through $U^\dagger {\bf z}$, 
and their c.c.s through $U {\bf z}^*$, 
$U \partial_{{\bf z}}$, etc. All quantities ${\mathfrak  H}, G, \*$, 
are scalar dot products.  Applications of such generating functions
are detailed in \cite{cuz}.

Several constructions based on the potentials introduced here will be
presented in future publications.

\noindent{\Large{\bf Acknowledgments}}\newline 
This work was supported in part by the US Department of Energy, 
Division of High Energy Physics, Contract W-31-109-ENG-38, and the NSF Award 
0073390.


\begin{thebibliography} {99} 
\bibitem{moyal} J Moyal, Proc Camb Phil Soc {\bf 45} (1949) 99-124
\bibitem{wigner} E Wigner, Phys Rev {\bf 40} (1932) 749-759
\bibitem{weyl} H Weyl, Z Phys {\bf 46} (1927) 1; also reviewed in H Weyl 
 (1931) {\em The Theory of Groups and Quantum Mechanics}, Dover, New York
\bibitem{bayen} F Bayen, M Flato, C Fronsdal, A Lichnerowicz, and D 
  Sternheimer, Ann Phys {\bf 111} (1978) 61; {\em ibid.}~111  
\bibitem{reviews}   C Zachos, hep-th/0110114;\\
 M Hillery, R O'Connell, M Scully, and E Wigner, Phys 
 Repts {\bf 106} (1984) 121;\\  
 H-W Lee, Phys Repts {\bf 259} (1995) 147;\\
 N Balasz and B Jennings, Phys Repts {\bf 104} (1984) 347;\\
 R Littlejohn, Phys Repts {\bf 138} (1986) 193;\\
 M Berry, Philos Trans R Soc London {\bf A287} (1977) 237;\\  
 M Gadella, Fortschr Phys {\bf 43} (1995) 3, 229-264;\\   
 L Cohen, {\it Time-Frequency Analysis} 
        (Prentice Hall PTR, Englewood Cliffs,  1995);  \\ 
 F Berezin, Sov Phys Usp {\bf 23} (1980) 763-787  
\bibitem{sylvester} J Sylvester, Johns Hopkins University Circulars {\bf I}
  (1882) 241-242; ibid. {\bf II} (1883) 46; ibid. {\bf III} (1884) 7-9. 
  Summarized in {\em The Collected Mathematics Papers of James Joseph Sylvester}
  (Cambridge University Press, 1909) v. III
\bibitem{fft} T Santhanam and A Tekumalla, Found Phys {\bf 6} (1976) 583-587;\\
  E Floratos and G Leontaris, Phys Lett {\bf B412} (1997) 35-41
\bibitem{fz} D Fairlie and C Zachos, Phys Lett {\bf B224} (1989) 101;\\
  E Floratos, Phys Lett {\bf B228} (1989) 335-340
\bibitem{superintegrable} R Chatterjee, 
  Lett Math Phys {\bf 36} (1996) 117-126;\\
  Y Nambu, Phys Rev {\bf D7} (1973) 2405-2441;\\
  N Mukunda and E Sudarshan,  Phys Rev {\bf D13} (1976) 2846-2850;\\ 
  C Gonera and Y Nutku, Phys Lett {\bf A285} (2001) 301-306 
\bibitem{groen} H Groenewold, Physica {\bf 12} (1946) 405-460
\bibitem{kim} D Han, Y Kim, and M Noz, Phys Rev {\bf A40} (1989) 902-912 
\bibitem{cfz} T Curtright, D Fairlie, and C Zachos, Phys Rev {\bf D58} (1998) 
  025002
\bibitem{cuz} T Curtright, T Uematsu, and C Zachos, J Math Phys {\bf 42} 
  (2001) 2396-2415 [hep-th/0011137] 
\bibitem{dodonov} V Dodonov and V Man'ko, Physica {\bf 137A} (1986) 306-316
\end{thebibliography}
\end{document}